\documentclass[aps,pra,reprint,onecolumn,groupedaddress]{revtex4-2}

\usepackage{amssymb,amsmath,mathtools,stmaryrd}
\usepackage[capitalize]{cleveref}

\begin{document}

%\preprint{}

\title{Comment on ``Physical significance of artificial numerical noise in direct numerical simulation of turbulence''}


\author{Ryan M. McMullen}
\email[]{rmmcmul@sandia.gov}
%\thanks{}
\author{Michael A. Gallis}
\affiliation{Engineering Sciences Center, Sandia National Laboratories, P.O. Box 5800, Albuquerque, New Mexico 87185-0840, USA}

\author{Ishan Srivastava}
% \email[]{}
%\thanks{}
\author{Andrew J. Nonaka}
\author{John B. Bell}
\email[]{jbbell@lbl.gov}
\affiliation{Lawrence Berkeley National Laboratory, 1 Cyclotron Rd., Berkeley, California 94720, USA}

\author{Alejandro L. Garcia}
% \email[]{}
%\thanks{}
\affiliation{Department of Physics and Astronomy, San Jose State University, 1 Washington Square, San Jose, California 95192, USA}


\date{\today}

\begin{abstract}
Recently, Liao and Qin [J. Fluid Mech. \textbf{1008}, R2 (2025)] claimed that numerical noise in direct numerical simulation of turbulence using the deterministic Navier-Stokes equations is ``approximately equivalent'' to the physical noise arising from random molecular motion (thermal fluctuations).
% This claim is false. 
We show here that it this claim not supported by their results and that it contradicts other results in the literature.
Furthermore, we demonstrate that the numerical implementation of thermal fluctuations in their so-called ``clean numerical simulations'' is incorrect.
\end{abstract}

%\keywords{}

\maketitle

\section{Introduction}

Hydrodynamic turbulence is usually modeled using the deterministic Navier-Stokes (NS) equations, which treat the fluid as a continuum, thereby ignoring its molecular nature.
Scaling arguments~\cite{corrsin1959,tennekes1972,moser2006} suggest that the NS equations should accurately describe all scales of turbulence, including dissipation-range scales, since the Kolmogorov length and time scales are typically much larger than the corresponding molecular scales, e.g., the mean free path and mean collision time in gases.
However, these arguments do not consider the random fluctuations arising from thermal motion of the fluid's constituent molecules.
Although these thermal fluctuations are typically only considered to be important in mesoscale fluid systems, the precipitous drop-off of the turbulent kinetic energy spectrum in the dissipation range led to the prediction that thermal fluctuations could become comparable to turbulent fluctuations beginning at scales not much smaller than the Kolmogorov scale, even when it is much larger than the molecular scale~\cite{Betchov_1957,bandakPRE}.
While measuring dissipation-range statistics remains a considerable experimental challenge, this prediction has been independently verified by three completely different numerical techniques that account for thermal fluctuations, namely fluctuating hydrodynamics (FHD)~\cite{bellJFM}, direct simulation Monte Carlo (DSMC)~\cite{mcmullenPRL}, and molecular dynamics (MD)~\cite{komatsu2014}.
Notably, comparing these simulation results with direct numerical simulations (DNS) of the deterministic NS equations reveals that dissipation-range scales are dramatically modified by thermal fluctuations, and, consequently, that the deterministic NS equations do not accurately describe these scales.

Despite this, the recent paper by Liao and Qin~\cite{LQ}, hereafter referred to as LQ, claims that numerical noise due to truncation and round-off errors in DNS of the deterministic NS equations is ``approximately equivalent'' to the physical noise from thermal fluctuations.
Here, we argue that this claim is not supported by the evidence presented in LQ and that it contradicts the extant literature on thermal fluctuations in turbulence.
We further show that their numerical implementation of thermal fluctuations is incorrect.

\section{Thermal fluctuations in the turbulent energy spectrum \label{sec:spec}}

An essential feature of thermal fluctuations is that their covariance, and hence spectrum, is completely determined by equilibrium statistical mechanics~\cite{lifshitz2013}, which makes their signature unambiguous.
The energy spectra of thermal velocity fluctuations in two and three dimensions are respectively given by
\begin{align}
    & & & & E_\mathrm{th}^\mathrm{2D}(k) &= \frac{k_\mathrm{B} T}{2 \pi \rho} k, & E_\mathrm{th}^\mathrm{3D}(k) &= \frac{3 k_\mathrm{B} T}{4 \pi^2 \rho} k^2, & & & &
    \label{eq:Eth}
\end{align}
where $k_\mathrm{B}$ is the Boltzmann constant, $T$ is the temperature, and $\rho$ is the density.
Simulations of three-dimensional turbulence using DSMC~\cite{mcmullenPRL} and FHD~\cite{bellJFM} have shown that thermal fluctuations dominate the dissipation range of the turbulent energy spectrum, beginning at a crossover wavenumber $k_c\eta \sim 3$, where $\eta$ is the Kolmogorov scale.
Simulations of two-dimensional turbulence have shown an analogous crossover occurs at $k_c\eta_\varOmega \sim 3$, where $\eta_\varOmega$ is the enstrophy dissipation scale~\cite{ma2024}.
For wavenumbers greater than $k_c$, the energy spectrum coincides with the equilibrium thermal-fluctuation spectrum given by \cref{eq:Eth}.
Remarkably, the value of $k_c\eta$ depends very weakly (logarithmically at leading order) on the dimensionless thermal noise strength $\varTheta_\eta = k_\mathrm{B}T/\rho u_\eta^2 \eta^3$~\cite{bellJFM}, where $u_\eta$ is the Kolmogorov velocity scale, meaning that thermal fluctuations should dominate beginning at a scale comparable to the the Kolmogorov scale over a broad range of conditions.

LQ presents simulations of two-dimensional turbulent Kolmogorov flow. 
Therefore, a conclusive and straightforward way to verify the claim that numerical noise in DNS of the deterministic NS equations is approximately equivalent to thermal noise would be to show that the numerical noise produces velocity fluctuations having a spectrum consistent with $E_\mathrm{th}^\mathrm{2D}$ given in \cref{eq:Eth}.
Indeed, because the maximum wavenumber in the highest-resolution simulations of LQ is $k_\mathrm{max}\eta_\varOmega \approx 18$, one would expect the energy spectrum to coincide with $E_\mathrm{th}^\mathrm{2D}$ over the majority of a decade in wavenumber.
However, the spectra shown in, e.g., their Fig. 4(a), show no such evidence of thermal fluctuations.

LQ also present so-called ``clean numerical simulations'' (CNS), which are supposedly devoid of the kinds of numerical noise present in DNS, but to which they introduce a random velocity field that they claim mimics thermal fluctuations.
As we will show in Sec.~\ref{sec:num}, their procedure for doing so is incorrect, but for now we note that the CNS spectra agree with the DNS spectra.
One may then ask why even the simulations that explicitly include random velocity fluctuations do not exhibit a thermal-fluctuation-dominated range consistent with \cref{eq:Eth}.
We suspect the reason for this is that the amplitude of the random velocity field in the CNS of LQ is extremely small, as we demonstrate next.

The random field is taken to be Gaussian white noise with zero mean and standard deviation $\sigma = 10^{-10}$.
The nondimensional thermal velocity fluctuations thus have variance given by~\cite{lifshitz2013}
\begin{equation}
    \left\langle u_\mathrm{th}^2 \right\rangle = \frac{k_\mathrm{B}T}{\rho U^2 V} = \sigma^2,
    \label{eq:uth}
\end{equation}
where $U = \sqrt{\chi L/2\pi}$ is the velocity scale, $L$ is the domain length scale, and $\chi$ is the forcing amplitude in the Kolmogorov flow. 
$V = \Delta^2 L_z$ is the volume of a computational cell, $\Delta = L/N$ is the grid spacing with $N$ cells in each direction, and $L_z$ is the implicit out-of-plane domain size.
One can thus calculate the value of $L_z$ required to realize thermal fluctuations having variance $\sigma^2$.
Doing so yields
\begin{equation}
    L_z = \frac{k_\mathrm{B}T N^2}{4\pi^2 \rho \nu^2 \sigma^2 \mathrm{Re}^2},
    \label{eq:Lz}
\end{equation}
where $\mathrm{Re} = \chi^{1/2}(L/2\pi)^{3/2}/\nu$ is the Reynolds number, and $\nu$ is the kinematic viscosity.
LQ considers a Reynolds number $\mathrm{Re} = 2000$, $N = 1024$, and water at room temperature, for which $T = 293$~K, $\rho = 998$~$\text{kg m}^{-3}$, and $\nu = 1.00 \times 10^{-6}$~$\text{m}^2\text{ s}^{-1}$.
Substituting these values into \cref{eq:Lz} gives $L_z \approx 2.7 \times 10^6$~m (approximately 40\% of Earth's radius).
% Similarly, for air at room temperature, $L_z = 9.8 \times 10^6$~m (approximately 150\% of Earth's radius).
To further contextualize this, we can define the two-dimensional analogue of $\varTheta_\eta$ as $\varTheta_\varOmega=k_\mathrm{B}T/\rho u_\varOmega^2 \eta_\varOmega^2 L_z$, where $u_\varOmega = \nu/\eta_\varOmega$. 
Using \cref{eq:Lz} gives $\varTheta_\varOmega = (2 \pi \sigma \mathrm{Re} / N)^2 \approx 1.5 \times 10^{-18}$. 
By contrast, typical values of $\varTheta_\eta$ for terrestrial turbulent flows are $\sim 10^{-9}$--$10^{-6}$~\cite{bandakPRE}.

We conclude that the thermal noise strength used in LQ is many orders of magnitude smaller than what is representative of any physically realizable turbulent flow.
Had a more realistic value of $\sigma$ been used, the high-wavenumber portion of the CNS energy spectrum would presumably coincide with $E_\mathrm{th}^\mathrm{2D}$, whereas the corresponding DNS spectrum would not, which would in turn contradict the claim that numerical noise is equivalent to thermal noise. 
That claim is also contradicted by the results of the existing literature on thermal fluctuations in turbulence~\cite{Betchov_1957,bandakPRE,mcmullenPRL,bellJFM,ma2024}, all of which compared simulations that include thermal fluctuations to DNS of the NS equations.
In all cases, the energy spectra from the simulations including thermal noise are dramatically different than those from DNS for $k\eta \gtrsim 3$.
We note that LQ state this fact in their conclusions section yet make no attempt to reconcile it with their results.

\section{Numerical implementation of thermal fluctuations \label{sec:num}}

As mentioned in Sec.~\ref{sec:spec}, LQ attempt to incorporate thermal fluctuations in their CNS by adding a random velocity field $u_\mathrm{th}$ to the deterministic velocity field $u = -\partial_y \psi$, where $\psi$ is the streamfunction.
The modified streamfunction is given by
\begin{equation*}
    \psi^*(x,y,t) = \psi(x,y,t) - \int_0^y u_\mathrm{th} \, \mathrm{d}y,
\end{equation*}
which is then substituted into the equation of motion and integrated forward in time by $\Delta t$.
However, this procedure is incorrect for two reasons.

The first is that it does not respect the framework of FHD, which is the correct way to account for thermal fluctuations at the mesoscopic level.
Indeed, FHD has been justified from microscopic principles for both linearized~\cite{bixon1969,fox1970a,fox1970b} and nonlinear hydrodynamics~\cite{zubarev1983,espanol1998}.
Importantly, thermal fluctuations do not enter the equations of motion directly as a fluctuating velocity field.
Rather, they enter via a random stress tensor that is constrained to satisfy the fluctuation-dissipation relation~\cite{zarate2006fhd}.
This distinction is crucial because, while FHD successfully reproduces the the long-range correlations that are the hallmark of hydrodynamic fluctuations in nonequilibrium systems~\cite{zarate2006fhd}, the approach taken in LQ always yields spatially uncorrelated fluctuations corresponding to equilibrium. Specifically, if $\delta u_i$ and $\delta u_j$ are the $u$ velocity fluctuations about a stationary state at spatial locations $x_i$ and $x_j$, respectively, then $\langle \delta u_i \delta u_j \rangle = \langle \delta u_i \delta u_j \rangle_\mathrm{eq} = \langle u_\mathrm{th}^2 \rangle \delta_{ij}$, where $ \langle u_\mathrm{th}^2 \rangle$ is given by \cref{eq:uth}.
% As discussed in Sec.~\ref{sec:spec}, the dissipation range of the energy spectrum is dominated by equilibrium thermal fluctuations, so one

The second reason the numerical implementation of thermal fluctuations in LQ is incorrect is that their temporal integration scheme is not a valid method for numerically solving stochastic differential equations (SDEs).
A great deal of effort has been devoted to developing accurate temporal integrators for FHD (see, e.g., Ref.~\cite{delong2013}).
Here, we simply point out that even for the simplest Euler-Maruyama scheme~\cite{pavliotis2014}, the additive white noise term representing thermal fluctuations should have Wiener increments of $O(\sqrt{\Delta t})$, whereas in LQ they are $O(1)$.
Consequently, their scheme does not converge to the solution of the continuous SDE in the limit $\Delta t \rightarrow 0$.
% the random velocity field is wholly inconsistent with both specialized temporal integration schemes that have been developed for FHD (see, e.g., Delong et al., \emph{Phys. Rev. E} {87}, 2013, 033302) and standard numerical methods for solving stochastic differential equations, such as the simple Euler-Maruyama scheme.
For these reasons, the numerical method in LQ almost certainly produces incorrect results when random fluctuations are included.

\section{Conclusions}

We have argued here that the central claim of LQ -- that numerical noise in DNS is approximately equivalent to thermal fluctuations -- is not supported by their results. 
The claim would be verified if it could be shown that the statistics of numerical noise are consistent with those of thermal fluctuations, which are unambiguous.
However, their results do not show this.
For the corresponding CNS that explicitly (and erroneously) incorporate thermal fluctuations, we demonstrated that the noise variance used is at least 9 orders of magnitude smaller than what is representative of thermal fluctuations in most physically realizable turbulent flows, and we suspect that if a more realistic noise variance were used, the CNS results would contradict the central claim of LQ.
Further, the claim contradicts results in the literature on thermal fluctuations in turbulence~\cite{Betchov_1957,bandakPRE,mcmullenPRL,bellJFM,ma2024}, all of which show that accounting for thermal fluctuations dramatically modifies the energy spectrum relative to what DNS predicts for $k\eta \gtrsim 3$. 
Additionally, it was demonstrated that the method for simulating thermal fluctuations in the CNS of LQ has two major flaws. In particular, it does not produce the correct correlations out of equilibrium, and the temporal integration scheme is not a valid method for numerically solving SDEs.

In summary, the main claim of LQ is not supported by their own results, is at odds with other published work, and relies on incorrect numerical implementation of thermal fluctuations.
The purpose of this Comment article is thus to prevent the spread of the misconception that numerical and thermal noise are equivalent.

\begin{acknowledgments}
Sandia National Laboratories is a multimission laboratory managed and operated by National Technology \& Engineering Solutions of Sandia, LLC, a wholly owned subsidiary of Honeywell International Inc., for the U.S. Department of Energy’s National Nuclear Security Administration under contract DE-NA0003525.
This paper describes objective technical results and analysis. 
Any subjective views or opinions that might be expressed in the paper do not necessarily represent the views of the U.S. Department of Energy or the United States Government.
This work was supported by the U.S. Department of Energy (DOE), Office of Science, Office of Advanced Scientific Computing Research, Applied Mathematics Program under contract No. DE-AC02-05CH11231.
The authors thank G. L. Eyink for fruitful technical discussions.
\end{acknowledgments}

\bibliography{refs}

\end{document}